\documentclass[12pt]{iopart}
\usepackage[latin1]{inputenc}
\usepackage{graphicx,psfrag}
\usepackage[hypertex]{hyperref}

\newcommand{\average}[1]{\left\langle{#1}\right\rangle}
\newcommand{\E}{\rme}
\newcommand{\D}{\rmd}
\newcommand{\I}{\rmi}
\newcommand{\f}{\mathrm{f}}
\newcommand{\PP}{\mathcal{P}}
\newcommand{\tm}{{t_\mathrm{m}}}

\begin{document}

\letter{The distribution function of entropy flow in stochastic systems}
\author{A Imparato$^1$ and L Peliti$^2$}
\address{$^1$ Dipartimento di Fisica,  INFN-Sezione di Torino,
CNISM-Sezione di Torino\\ Politecnico di Torino, Corso Duca degli
Abruzzi 24, 10129 Torino (Italy)}
\address{$^2$ Dipartimento di Scienze Fisiche,
INFN-Sezione di Napoli, CNISM-Sezione di Napoli \\
Università ``Federico II'', Complesso Monte S. Angelo, 80126 Napoli
(Italy)} \eads{\mailto{alberto.imparato@polito.it},
\mailto{peliti@na.infn.it}}

\begin{abstract}
We obtain a simple direct derivation of the differential equation
governing the entropy flow probability distribution function of a
stochastic system first obtained by Lebowitz and Spohn. 
Its solution agrees well with the experimental results of
Tietz \etal [2006 {\it Phys. Rev. Lett.} {\bf 97} 050602]. 
A trajectory-sampling algorithm allowing to evaluate the entropy flow
distribution function is introduced and discussed. This algorithm turns out 
to be effective at finite times and in the case of 
time-dependent transition rates, and is successfully 
applied to an asymmetric simple exclusion process.
\end{abstract}

\vspace{0.5cm} \noindent{\it Keywords\/}: Dynamical processes
(Theory), Fluctuations (Theory)

\maketitle

The concept of entropy is usually associated with probability
distributions (ensembles) via Gibbs's formula. However, it has been
recently emphasized that in nonequilibrium systems one can
consistently define the entropy production along a single
trajectory~\cite{Crooks,Qian,Seifert}, and that the fluctuation
theorems~\cite{Crooks,Seifert,Evans,GC,Jarzynski,Kurchan,LS,Maes,Gaspard,Seifert2,noi4}
can be interpreted as giving connections between the probability of
entropy-generating trajectories with respect to that of
entropy-annihilating trajectories. Since the entropy flow
along a given trajectory is experimentally accessible~(see, e.g.,
\cite{Seif_exp}), it is of some interest to investigate the
distribution function of the entropy production and flow in a
nonequilibrium system. The distribution of entropy flow in a
stochastic system satisfies a differential equation derived by
Lebowitz and Spohn in~\cite{LS}. They made use of this equation to
investigate the large-deviation function of entropy production in
the long-time limit, thus obtaining a form of the Gallavotti-Cohen
symmetry~\cite{GC}. In the present paper, we provide a new direct
derivation of this differential equation in a general nonequilibrium
system with discrete states, and show that its solution satisfies
more general fluctuation relations also at finite times. This
derivation highlights its connection with the entropy production
along a single trajectory. We exploit it to evaluate the entropy
flow in experiments on an optically driven defect center in
diamond~\cite{Seif_exp}. We suggest a practical computational scheme
to evaluate the distribution of entropy flow. We illustrate the
feasibility of the method by applying it to the simple asymmetric
exclusion process (ASEP)~\cite{SD}.

We consider a system with a discrete phase space, whose states are
denoted by $i=1, \dots, N$. We assume that the evolution of the
system is described by a Markovian stochastic dynamics
\begin{equation}
\frac{\D p_i(t)}{\D t} =\sum_{j\,(\neq i)} \left[W_{ij}(t)
p_j(t)-W_{ji}(t) p_i(t)\right], \label{master_p}
\end{equation}
where $p_i(t)$  is the probability that the system is found at state
$i$ at time $t$, and $W_{ij}(t)$ is the transition rate from the
state $j$ to the state $i$ at time $t$. We consider a generic path
$\omega$ defined by $\omega(t)=i_k$ iff $t_k\le t < t_{k+1}$, with
$k=0,1,\ldots,M$, with $t_{M+1}={t_\f}$, and define the
time-reversed path $\widetilde\omega$ by $\widetilde\omega(t)=i_k$
iff $t\in[\tilde t_{k+1},\tilde t_{k})$, where $\tilde
t=t_0+{t_\f}-t$. Let $Q(\omega)$ be defined in terms of the ratio
between the probability $\PP(\omega)$ of the forward path $\omega$
(conditioned by its initial state $i_0$) and the probability
$\widetilde \PP(\widetilde \omega)$ of the time-reversed path
$\widetilde \omega$, conditioned by \textit{its} initial state
$i_M\equiv i_{\f}$ and subject to the time-reversed protocol
$\widetilde W_{ij}(t)=W_{ij}(\tilde t)$~\cite{Crooks,Seifert,LS}:
\begin{equation}
 Q(\omega)=-\ln
    \left[\frac{\PP(\omega)}{\widetilde\PP(\widetilde\omega)}\right]
    =-\sum_{k=1}^M \ln \left[\frac{W_{i_{k}i_{k-1}}(t_k)}{W_{i_{k-1}
    i_{k}}(t_k)}\right].
\label{entrflow_def}
\end{equation}
We have assumed that, if $W_{ij}(t)>0$ at any time $t$, one also has
$W_{ji}(t)>0$. In order to interpret this quantity, let us consider
the Gibbs entropy of the system $S\mathrm(t)=-\sum_i p_i(t) \ln
p_i(t)$ (we take the Boltzmann constant $k_\mathrm{B}=1$
throughout). Using~\eref{master_p}, the time derivative of
$S_\mathrm{S}$ reads
\begin{equation}
 \frac{\D S}{\D t}=-\sum_{i \neq j} W_{ij} p_j \ln
   \left(\frac{W_{ji}p_i}{W_{ij}p_j}\right)-\sum_{i \neq j} W_{ij}
   p_j\ln\frac{W_{ij}}{W_{ji}}.
\end{equation}
By exploiting the relation $\ln x \le x-1$, one sees that the first
sum is non-negative, and can thus be interpreted as the entropy
production rate~\cite{LS,Gaspard}. The second sum defines the entropy
$S_{\f}$ which flows into the reservoir:
 $\D S_{\f}/\D t=\sum_{i \neq j} W_{ij} p_j\ln\left(W_{ij}/W_{ji}\right)$.
It is easy to verify that, if we define $\average{Q}_t$ as the
average of the quantity $Q(\omega)$, defined by~\eref{entrflow_def},
over all possible paths up to time $t$, the following equality
holds: $\D \average{Q}_t/\D t =\D S_{\f}/\D t$, and thus
$\average{Q}_{t_\f}  =\Delta S_{\f}=\int_{t_0}^{t_\f} \D t\; 
\D S_{\f}/\D t$.
Notice that if the detailed balance conditions holds for the
transition rates $W_{ij}(t)$, and the energy $E_i(t)$ is associated
to the state $i$ of the system, we have
$W_{ji}(t)/W_{ij}(t)=\exp\left\{
\left[E_i(t)-E_j(t)\right]/T\right\}$, and thus
$T\ln\left[W_{ji}(t)/W_{ij}(t)\right]$ represents the heat exchanged
with the reservoir in the jump from state $j$ to state $i$. Thus the
quantity $Q(\omega)$, defined by~\eref{entrflow_def}, is the entropy
which flows into the reservoir as the system evolves along the path
$\omega$ (see, e.g.,~\cite{Seifert}). We now introduce the joint
probability distribution $\phi_i(Q,t)$, that the system is found at
time $t$ in state $i$, having exchanged a total entropy flow $Q$
with the thermal bath. The entropy which flows into the reservoir as
a result of the jump of the system from state $j$ to state $i$ is
given by $\Delta s_{ij} = \log\left[W_{ji}(t)/W_{ij}(t)\right]$. In
a small time interval $\tau$ the variation of $\phi_i(Q,t)$ is given
by
\begin{eqnarray}
    &\phi_i(Q,t+\tau)\simeq \phi_i(Q,t)
    +\tau\sum_{j\,(\neq i)} W_{ij} \phi_j(Q-\Delta s_{ij}, t)-W_{ji}\phi_i(Q,t) \nonumber \\
    &\quad{}=\phi_i(Q,t)+\tau \sum_{j\,(\ne i)} \left\{W_{ij}
    \left[\sum_{n=0}^\infty\frac{\left(-\Delta
    s_{ij}\right)^n}{n!}\frac{\partial^n\phi_j(Q,t)}{\partial Q^n} \right]
     -W_{ji}\phi_i(Q,t)\right\},
\label{var_phi}
\end{eqnarray}
and thus we obtain the differential equation governing the time
evolution of the distribution function $\phi_i(Q,t)$:
\begin{equation}
    \frac{\partial \phi_i(Q,t)}{\partial t}=\sum_{j\,(\neq i)}
    \left\{ W_{ij} \left[\sum_{n=0}^\infty
    \frac{\left(-\Delta s_{ij}\right)^n}{n!}
    \frac{\partial^n\phi_j(Q,t)}{\partial Q^n} \right]
    -W_{ji}\phi_i(Q,t)\right\}.
\label{eq_phi}
\end{equation}
By introducing, for each $i$, the generating function
\begin{equation}
\psi_i(\lambda,t)=\int \D Q\; \E^{\lambda Q} \phi_i(Q,t),
\label{psi}
\end{equation}
and taking into account the expression of $\Delta s_{ij}$, we obtain
the master equation
\begin{eqnarray}
    \frac{\partial\psi_i(\lambda,t)}{\partial t}&=\sum_{j\,(\ne i)}
    \left[W_{ij} \left(\frac{W_{ji}}{W_{ij}}\right)^\lambda
    \psi_j(\lambda,t) -W_{ji}\psi_i(\lambda,t)\right]\nonumber\\
    &=\sum_j H_{ij}(\lambda)\,\psi_j(\lambda,t) . \label{master}
\end{eqnarray}
This last equation was first derived by Lebowitz and Spohn in~\cite{LS}.

We now introduce the total probability distribution  
$\phi(Q,t)=\sum_i\phi_i(Q,t)$, and the total generating function $\psi(\lambda,t)=\sum_i\psi(\lambda,t)$.
In the case of time-independent transition rates $W_{ij}$, or of
transition rates which depend periodically on the time, it can be
useful, in order to evaluate the distribution function $\phi(Q,t)$,
to introduce the large-deviation function. In the long-time limit,
the generating function $\psi(\lambda,t)$ is dominated by the
maximum eigenvalue $g(\lambda)$ of the matrix
$\mathsf{H}(\lambda)=\left(H_{ij}(\lambda)\right)$, which appears in
the master equation \eref{master}. Therefore, we have, for long
times $t$,
\begin{equation}
\psi(\lambda,t)\propto\exp\left[t \,g(\lambda)\right]. \label{gf}
\end{equation}
By using the last equation and inverting~\eref{psi}, one obtains the
probability distribution of the entropy flow in the long time limit
as
\begin{equation}
\phi(Q,t)=\int \frac{\D \lambda}{2\pi\I}\; \E^{-\lambda Q}
\psi(\lambda, t)
    \propto \E^{t\,g(\lambda^*)-\lambda^* Q}, \label{large_phi}
\end{equation}
where $\lambda^*$ is the saddle point value implicitly defined by
$\left.\partial g/\partial \lambda\right|_{\lambda^*}=Q/t$. If we
introduce the entropy flow per unit time $q=Q/t$, we obtain the
large-deviation function
\begin{equation}
    f(q)\equiv g(\lambda^*)-\lambda^* q
    =\lim_{t\rightarrow\infty} \frac 1 t \log \phi(Q,t).
\label{fq_def}
\end{equation}
Note that the functions $g(\lambda)$ and $f(q)$ are Legendre
transform of each other, and can be then interpreted in terms of
path thermodynamics: $g(\lambda)$ can be viewed as a path Gibbs free
energy, while $f(q)$ is the corresponding Helmholtz free energy.
This analogy was first pointed out in~\cite{felix,noi3} where work
probability distributions of systems driven out of equilibrium were
studied. The Gallavotti-Cohen theorem~\cite{GC}
$\phi(Q)/\phi(-Q)=\exp(Q)$ is immediately recovered by considering
that $\mathsf{H}(-\lambda+1)=\mathsf{H}^\mathsf{T}(\lambda)$, where
$\mathsf{H}^\mathsf{T}$ is the transpose of $\mathsf{H}$, and thus
has the same eigenvalues as $\mathsf{H}$ \cite{LS}.

Another remarkable fluctuation relation has been recently proposed
by Seifert~\cite{Seifert}. Since the path probability densities
$\PP(\omega)$ and $\widetilde\PP(\widetilde \omega)$ must be
normalized, one obtains from~\eref{entrflow_def}, for any distribution
of initial states $p_i^0$ and any distribution of final states
$p_i^\f$ the following fluctuation relation,
\begin{eqnarray}
    \average{
    \exp\left[Q(\omega)+\ln\left(\frac{p^\f_{i_\f}}{p^0_{i_0}}\right)\right]}
    &\equiv \sum_{i_0,i_\f}\int_{\omega({t_0})=i_0}^{\omega({t_\f})=i_\f}
     \mathcal{D}\omega\; p^0_{i_0}
     \PP(\omega) \,\E^{Q(\omega)}\, \frac{p^\f_{i_\f}}{p^0_{i_0}}\nonumber\\
    &= \sum_{i_0,i_\f}\int^{\tilde\omega({t_\f})=i_0}_{\tilde \omega({t_0})=i_\f}
\mathcal{D}\widetilde \omega\;
    p^{\f}_{i_\f}\widetilde\PP(\widetilde \omega) =1.
\label{seifert_rel}
\end{eqnarray}
Starting from~\eref{master}, this relation is recovered as follows.
Let $\psi_j^{i_0}(\lambda,t)$ be the solution of~\eref{master} with
the initial condition $\psi_j^{i_0}(\lambda,t{=}0)=\delta_{j,i_0}$,
and let $\bar \psi_j(\lambda,t)=\sum_{i_0}\psi^{i_0}_j(\lambda,t)$.
The function $\bar \psi_j(\lambda{=}1,t)=1, \, \forall t$, is the
solution of~\eref{master} with the initial condition $\bar
\psi_j(\lambda{=}1,t{=}0)=1$. Thus, \eref{seifert_rel} reads
\begin{eqnarray}
    \average{\exp\left[Q+\ln\left(p^{\f}_{i_\f}/p^0_{i_0}\right)\right]}&=\sum_{i_0,j}
    p^0_{i_0}
    \psi_j^{i_0}(\lambda{=}1,t)\,\exp\left[\ln\left(p_j(t)/p^0_{i_0}\right)\right]\nonumber\\
    &=\sum_j p_j(t)  \bar \psi_j(\lambda{=}1,t)=1,
\end{eqnarray}
 for any value of $t$. This result is a generalization of equation
(2.20) of~\cite{LS} and has been recently applied to a specific case
in~\cite{Harris}.

If the system is characterized by a small number of states, one can
explicitly solve the equations~\eref{master}, and thus obtain the
total generating function $\psi(\lambda,t)=\sum_i
\psi_i(\lambda,t)=\average{\E^{\lambda Q}}.$

\begin{figure}[h]
\center
\psfrag{q}[ct][ct][1.]{$-q$}
\includegraphics[width=8cm]{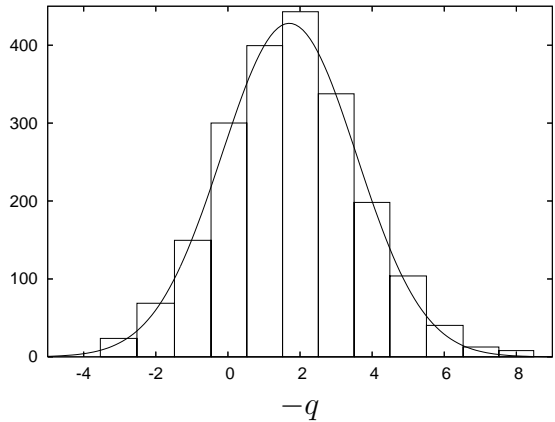}
\caption{Entropy production of  a two-state system: a defect center
in diamond is optically driven from a fluorescent to a
non-fluorescent state. The experimental histogram is taken
from~\cite{Seif_exp}, the full line corresponds to the entropy flow distribution
function $\phi(Q,t=20\,\tm)$ as obtained from
(\ref{eq_phi}--\ref{large_phi}). Note that the definition of $q$ in \cite{Seif_exp} is the negative of ours.} \label{Seif_fig}
\end{figure}

As an example, we consider an optically driven defect center in
diamond, which can be viewed as a  two-state system \cite{Seif_exp}.
If excited by a red light laser, the defect exhibits  fluorescence.
By superimposing a green light laser, the rate of transition from
the non-fluorescent to the fluorescent state ($W_+$ in the following)
and from the
fluorescent to the non-fluorescent state ($W_-$), turn out to depend
linearly on the green and red light intensity, respectively.
In~\cite{Seif_exp}, the experimental set-up was such that the rate
$W_-=(21.8\,
\mathrm{ms})^{-1}$ was kept
constant, and the rate $W_+$  was modulated according to the sinusoidal
function $W_+(t) = W_0 \left[1 +\gamma \sin(2\pi t/\tm)\right]$,
with $W_0=(15.6\, \mathrm{ms})^{-1}$, $\gamma=0.46$,  $\tm=50\,\mathrm{ms}$. In the same reference,
the histogram of the entropy flux, as defined
by~\eref{entrflow_def}, was measured, over 2000 trajectories of
time-length $20\,\tm$. By solving~\eref{master} for this system, we
obtain the generating function $\psi(\lambda,t)$, and
using~\eref{large_phi} we obtain the probability distribution of the
entropy flow $\phi(Q,t=20\,\tm)$, which is found to agree very well
with the measured histogram, see~\fref{Seif_fig}.

However this direct approach becomes rapidly impracticable, as the
system phase space size increases. On the other hand, the evaluation
of $\phi(Q,t)$ or $\psi(\lambda,t)$ by direct simulation of the
stochastic process described by the master equation \eref{master_p}
is a highly difficult task. Similarly to what happens when one tries
to evaluate free energies differences by using the Jarzynski
equality~\cite{Jarzynski}, dealing with the probability distribution
of the work of a driven system (see, e.g.,~\cite{felix,noi3}), the
most relevant contributions to the average $\average{\exp(\lambda
Q)}$ often come from the tails of the distribution of $\phi(Q,t)$.

In ref.~\cite{GKP}, the authors proposed a procedure to evaluate the 
large deviation function $g(\lambda)$ based on biased dynamics and on 
the parallel evolution of system clones.
Here we discuss how~\eref{master} leads to an alternative  
scheme for the evaluation of the fluctuations in the entropy flow.
Differently from what discussed in ref.~\cite{GKP}, such a scheme 
allows one
 to evaluate correctly the function
$\psi(\lambda,t)=\average{\exp(\lambda Q)}$ at any time, and not only in the long time limit, and turns out to be effective also in the case of time--dependent transition rates.
In the following we adapt to our problem
the concept of weighted trajectory ensemble, introduced by Sun in
\cite{Sun}  and extended by Oberhofer \etal\ in~\cite{ODG}, where
it was successfully applied to the Jarzynski equality. The idea is
to sample trajectories in the weighted ensemble
$\PP(\omega_t)\exp\left[\lambda Q(\omega_t)\right]$ rather than
successions of single states in the unbiased ensemble. Since
$\psi(\lambda,t)=\int \mathcal{D}\omega_t\; \PP(\omega_t)\,
\E^{\lambda Q(\omega_t)}$, we have
\begin{equation}
\frac{\partial\psi(\lambda,t)}{\partial \lambda}
=\average{Q}_\lambda \psi(\lambda,t), \label{psidel}
\end{equation}
where $\average{\dots}_\lambda$ is the average in the weighted
ensemble $\PP(\omega_t) \exp\left[\lambda
Q(\omega_t)\right]\big/\psi(\lambda,t)$, where
$\psi(\lambda,t)=Z_\lambda$ is the ``partition function'' of this
ensemble, which will be called the ``$\lambda$-ensemble'' in the
following. The solution of~\eref{psidel} thus reads
\begin{equation}
\psi(\lambda,t)=\exp\left[\int_0^\lambda\D \lambda'
\average{Q}_{\lambda'}\right]. \label{psi_sun}
\end{equation}

While Sun~\cite{Sun} proposes a Montecarlo procedure for the
sampling of trajectories, so as to obtain the average
$\average{Q}_\lambda$ in the weighted ensemble, in the present paper
we propose a different procedure which generates trajectories in a
suitable entropy-flow weighted ensemble. The direct simulation of
trajectories in the $\lambda$-ensemble is hindered by the fact that
one should already know the exact expression of the ensemble
partition function, i.e., the function $\psi(\lambda,t)$, which is
the unknown quantity at issue.

To avoid the problem of the direct evaluation of $\psi(\lambda,t)$,
following~\cite{ODG}, we introduce a generic functional of the paths
$\Pi(\omega)$, and write
\begin{equation}
\average{Q}_\lambda=\frac{\int \mathcal{D}\omega_t\;
(Q(\omega_t)/\Pi(\omega_t))\Pi(\omega_t) \PP(\omega_t)\E^{\lambda
Q(\omega_t)}} {\int \mathcal{D}\omega_t\;
(1/\Pi(\omega_t))\Pi(\omega_t) \PP(\omega_t)\E^{\lambda Q(\omega_t)}
}
=\frac{\average{Q/\Pi}_{\lambda,\Pi}}{\average{1/\Pi}_{\lambda,\Pi}},
\label{qlp}
\end{equation}
where $\average{\dots}_{\lambda,\Pi}$ indicates the average in the
new $\PP(\omega_t) \Pi(\omega_t)\exp\left[\lambda
Q(\omega_t)\right]$ ensemble (the $(\lambda,\Pi)$-ensemble in the
following).

In order to simplify the following discussion
we consider a stochastic dynamics with a discrete small time scale
$\tau$, such that the jumps between states take place at discrete
times $t_k=k \tau$. Within this discrete scheme, we want to write
the probability of a path in the $(\lambda,\Pi)$-ensemble.

The probability of a given path $\omega$ reads
\begin{equation}
\PP(\omega)=K_{i_N,i_{N-1}} K_{i_{N-1},i_{N-2}}\cdots K_{i_1,i_{0}}p^0_{i_0},
\end{equation}
where the transition probabilities $K_{ij}$ are defined as
$K_{ij}=\tau W_{ij}$, and $K_{ii}=1-\sum_{j (\ne i)}K_{ji}$. We
now define the new transition probabilities $\widetilde K_{ij}=\tau
W_{ij} \left(W_{ji}/W_{ij}\right)^\lambda$, and $\widetilde
K_{ii}=1-\sum_{j (\ne i)}\widetilde K_{ji}$, and choose the
functional $\Pi(\omega)$, such that
\begin{equation}
\Pi(\omega)=\prod_{k=1}^{M}\Pi_{i_k,i_{k-1}}(t_k),
\end{equation}
with
\begin{equation}
\Pi_{ij}(t)= \cases{
    1,
    & if $i\ne j$ ;\cr
   \widetilde K_{jj}(t)/ K_{jj}(t),  & if
   $i=j$.}
\label{pij}
\end{equation}
Recalling the definition of $Q(\omega)$, \eref{entrflow_def}, we
obtain that the probability in the $(\lambda,\Pi)$-ensemble is given
by
\begin{equation}
\PP(\omega)\Pi(\omega) \exp\left[\lambda Q(\omega)\right]=
\widetilde K_{i_N,i_{N-1}}\widetilde  K_{i_{N-1},i_{N-2}}\cdots
\widetilde K_{i_1,i_{0}}p^0_{i_0}, \label{pp_pi_q}
\end{equation}
and thus $\average{Q}_\lambda$ can be evaluated by using~\eref{qlp},
for the particular choice of $\Pi$, as given by \eref{pij}. Note
that \eref{pp_pi_q} implies that the one can generate a trajectory
in the $(\lambda,\Pi)$-ensemble by simply simulating the process with
the $\widetilde K_{ij}(t)$ transition probabilities. Furthermore, the algorithm here described can be used to evaluate the generating function (\ref{psi_sun}) also in the case of time--dependent transition rates.
This issue will be addressed in a forthcoming paper.

\begin{figure}[h]
\center \psfrag{Gamma}[ct][ct][1.]{$g$ }
\psfrag{l}[ct][ct][1.]{$\lambda$ }
\includegraphics[width=8cm]{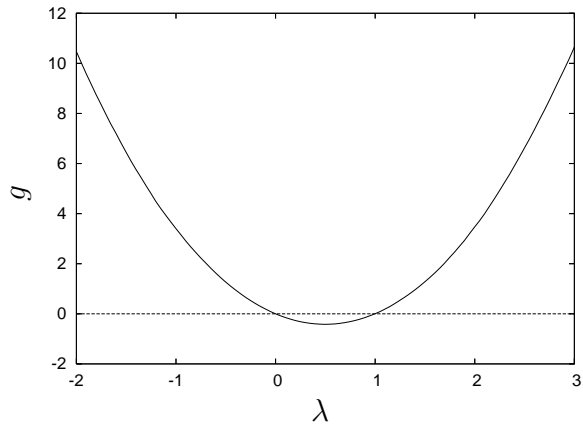}
\caption{Plot of $g(\lambda)$ as obtained by combining~\eref{gf} and
\eref{psi_sun}, for the ASEP model. The function $g(\lambda)$
vanishes for $\lambda=0,1$ which corresponds to the normalization
condition and to Seifert's fluctuation relation \eref{seifert_rel},
respectively.} \label{fig1}
\end{figure}

\begin{figure}[h]
\center
\psfrag{P}[ct][ct][1.]{$\phi$}
\psfrag{f}[lb][lb][1.][-90]{$f$}
\psfrag{q}[ct][ct][1.]{$q$}
\includegraphics[width=8cm]{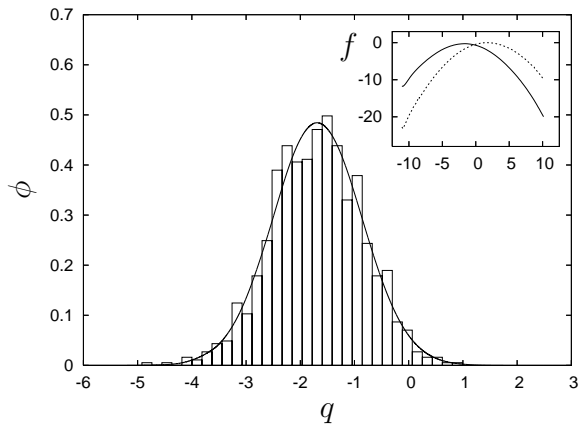}
\caption{Histogram of the entropy flow per time unit $q$,
corresponding to 1000 unbiased trajectories of the ASEP model. Full line: probability
distribution function $\phi(q,{t_\f})$, obtained with the trajectory
sampling algorithm. Inset: full line, large-deviation function $f$
as a function of the entropy per time unit $q$, as defined
by~\eref{fq_def}; dotted line: plot of
$f(q)+q$.} \label{fig2}
\end{figure}

\begin{figure}[h]
\center \psfrag{j}[ct][ct][1.]{$J$} \psfrag{r}[ct][ct][1.]{$\rho$}
\psfrag{l}[ct][ct][1.]{$\lambda$}
\includegraphics[width=8cm]{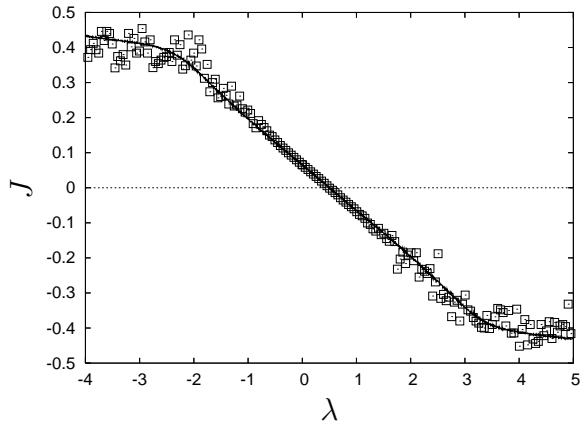}
\caption{Average particle current $J$ in the weighted trajectory
$\lambda$-ensemble (full line) and in the $(\lambda,\Pi)$-ensemble (squares), as a
function of $\lambda$. The current vanishes at $\lambda=1/2$, and
becomes negative for larger values of $\lambda$.} \label{fig5}
\end{figure}

In order to give an example of the application of the approach above
described, we now consider a nonequilibrium system characterized by
a large phase space, and evolving according to
a stochastic dynamics. Instead of considering systems driven out of
equilibrium by manipulation as in~\cite{noi3, noi}, we study here a
steady-state non-equilibrium system, namely the asymmetric simple
exclusion process (ASEP)~\cite{SD,ED}: it consists in a
one-dimensional lattice gas on a lattice of $L$ sites. Each site of
the model is either empty or occupied by at most one particle. Each
particle can jump into an empty nearest neighbor site with
transition rates per time unit $W_+$ (rightward) and $W_-$
(leftward). The system is kept in a out-of-equilibrium steady state
since its first and last site are in contact with two particle
reservoirs, at densities $\rho_A$ and $\rho_B$ respectively. By
taking $\rho_A>\rho_B$ and $W_+>W_-$, one observes a net particle
current from the left to the right reservoir. We choose the
following values of the parameters: $L=100$, $W_+=1$, $W_-=0.75$,
$\rho_A=0.75$, $\rho_B=0.25$, which correspond to the maximum
current phase~\cite{SD}.

A direct evaluation of the function $g(\lambda)$ by solving the
$2^{100}$ equations \eref{master} is of course impossible. We thus
apply our trajectory simulation approach to the ASEP model. We
consider trajectories of time length ${t_\f}=5$, with the elementary
time step $\tau=0.01$: at each time the transition probability
between two states is given by the transition matrix $\widetilde
K_{ij}(t)$. For each value of $\lambda$ we generate
$\mathcal{N}=1000$ sample trajectories and calculate the entropy
flow $Q$, as defined by~\eref{entrflow_def}, for each trajectory.
Then, for the given value of $\lambda$, by averaging over the
$\mathcal{N}$ trajectories, we compute the quantity
$\average{Q}_\lambda$ using~\eref{qlp}. Finally, by combining
\eref{gf} and~\eref{psi_sun}, we obtain the function $g(\lambda)$
which governs the long time behavior of $\psi(\lambda,t)$. This
function is plotted in~\fref{fig1}. It can be seen that it vanishes
at $\lambda=0,1$ and is symmetric with respect to $\lambda=1/2$. The
fact that $g(0)=0$ corresponds trivially to the normalization
condition over all the possible trajectories. On the other hand, the
fact that the function $g$ vanishes at $\lambda=1$, is a non trivial
result, and corresponds to Seifert's fluctuation theorem
\eref{seifert_rel}. The symmetry around $\lambda=1/2$ corresponds to
the Gallavotti-Cohen fluctuation relation. We are now able to
calculate the large-deviation function $f(q)$ defined
in~\eref{fq_def}, which is plotted in the inset of~\fref{fig2},
together with the expression $f(q)+q$, which exhibits the symmetry
required by the Gallavotti-Cohen relation. We check as follows that
the quantity $f(q)$ actually gives the entropy distribution function
$\phi(q,t)\propto \exp\left[t\, f(q)\right]$ for the present model.
We simulate the unbiased diffusion process (i.e., we use the
transition matrix $K_{ij}$), and measure the entropy flow along 1000
trajectories. We then plot the histogram of the measured entropy
flow per time unit, together with the function $ \exp\left[t\,
f(q)\right]$, see~\fref{fig2}. The agreement between the histogram
and the predicted entropy distribution $\phi(q,t)$ is excellent.
Note that it is hopeless to verify the Gallavotti-Cohen and/or the
Seifert relation \eref{seifert_rel} by direct inspection of the
histogram, since no single point of the histogram lies in the range
$q>1$, where the function $f(q)+q$ has its maximum. This is the same
problem that one faces when trying to exploit the Jarzynski equality
to evaluate the free energy of simple microscopic
systems~\cite{noi3}.

Finally, we consider the dynamic states which contribute most to the
function $g$ for each value of $\lambda$. We measure the current
$J(\lambda)$ of particles  that, in the steady state of weighted
ensembles, jumps to the right (positive current) or to the left
(negative current), in the unit time. This quantity can be easily
measured as biased trajectories are generated using the probability
$\widetilde K_{ij}$. The current $J$, as measured in both the
$\lambda$ and $(\lambda,\Pi)$-ensemble, is plotted in~\fref{fig5},
as a function of $\lambda$. It can be clearly seen that, as
$\lambda$ goes away from the origin (either in the negative or in
the positive direction), the current differs more and more from its
value in the unbiased dynamics ($\lambda=0$). Notice that the
current vanishes at $\lambda=1/2$, and that, for larger values of
$\lambda$, $J$ becomes negative, i.e., the net motion of the
particles is opposite to the density gradient. Thus, the parameter
$\lambda$, which is the thermodynamic conjugate of $q$, selects
dynamical trajectories in the same way as an external field selects
states in an ordinary statistical ensemble. Large values of
$|\lambda|$ select trajectories which are highly unlikely to appear
in an unbiased system. This corroborates the interpretation of the large deviation function as a kind of free energy in the path thermodynamics \cite{felix,noi3}.

We have thus shown how it is possible to evaluate consistently the
probability distribution function of the entropy flow in
nonequilibrium systems, by solving the differential
equation~\eref{master} via the simulation scheme for the
($\lambda,\Pi$)-ensemble. In this way the properties of the
trajectories which contribute most to the entropy flow in the regime
of interest can also be evaluated.

A.I. is grateful to M. Gianfelice for long and interesting discussions, and to 
C. Mejia-Monasterio for important remarks.

\end{document}